\documentclass[opinion]{sigirforum}

\def\pubissue{Vol. 57 No. 2 -- December 2023}

\usepackage{amsmath}
\usepackage{pifont}
\newcommand{\cmark}{\ding{52}}%
\newcommand{\xmark}{\ding{55}}%

\begin{document}
\title{Large Search Model: Redefining Search Stack \\ in the Era of LLMs}

\author{~\textbf{Liang Wang}\thanks{Equal contribution. Correspondence to \{wangliang,nanya,fuwei\}@microsoft.com},~\textbf{Nan Yang}$^*$,~\textbf{Xiaolong Huang},\\
~\textbf{Linjun Yang},~\textbf{Rangan Majumder},~\textbf{Furu Wei}\\
Microsoft Corporation \\}

\maketitle 
\begin{abstract}
Modern search engines are built on a stack of different components,
including query understanding, retrieval, multi-stage ranking, and question answering, among others.
These components are often optimized and deployed independently.
In this paper,
we introduce a novel conceptual framework called \emph{large search model},
which redefines the conventional search stack by unifying search tasks with one large language model (LLM).
All tasks are formulated as autoregressive text generation problems,
allowing for the customization of tasks through the use of natural language prompts.
This proposed framework capitalizes on the strong language understanding and reasoning capabilities of LLMs,
offering the potential to enhance search result quality while simultaneously simplifying the existing cumbersome search stack.
To substantiate the feasibility of this framework,
we present a series of proof-of-concept experiments
and discuss the potential challenges associated with implementing this approach within real-world search systems.
\end{abstract}

\section{Introduction}

With the ever-increasing amount of information on the web,
search engines have emerged as an indispensable tool for finding and accessing information.
Several decades of research efforts in the fields of information retrieval (IR), machine learning, and computational advertising
have culminated in the continuous evolution and commercial success of search engines like Google and Bing.
Although current systems still provide the traditional ten blue links as the primary search results,
the Search Engine Result Page (SERP) often contains various other valuable information,
including direct answers ~\citep{Potthast2021TheDO}, featured snippets, knowledge panels, related queries, and multimedia content, etc.
Recently,
the popularization of ChatGPT ~\footnote{\url{https://openai.com/blog/chatgpt}} has led to the emergence of conversational search interfaces,
where LLMs generate responses to user queries with retrieval augmentation.

To support the aforementioned user information needs,
modern search engines are built on a stack of different components,
including query intent understanding, embedding-based and lexical matching-based retrieval,
multi-stage ranking, question answering, summarization, etc.
Typically,
these components are designed and optimized independently,
with the standard practice of fine-tuning pre-trained language models
like BERT ~\citep{Devlin2019BERTPO} or T5 ~\citep{raffel2020exploring} on task-specific datasets.
To handle hundreds of thousands of queries per second,
knowledge distillation is also widely used to reduce the computational cost of inference.
However,
this results in a cumbersome search stack that is difficult to maintain.
Furthermore,
the quality of search results for long-tailed and complex information needs are still far from satisfactory.
A unified modeling framework that offers flexible interfaces and improved generalization would be a more desirable solution.

Large language models (LLMs) that leverage self-supervised pre-training and scaling laws ~\citep{Kaplan2020ScalingLF}
have emerged as general-purpose interfaces for natural language understanding and generation.
Cutting-edge LLMs like GPT-4 ~\citep{OpenAI2023GPT4TR} and LLaMA ~\citep{Touvron2023LLaMAOA}
are trained on web-scale corpora and demonstrate remarkable zero-shot and few-shot learning capabilities.
In fact,
they even surpass human performance on a variety of professional exams ~\citep{OpenAI2023GPT4TR}.
These desirable features position LLMs as a promising option for the unified modeling of search tasks.

In this paper,
we propose a conceptual framework denoted as \emph{large search model} that
reimagines the conventional search stack from an LLM perspective.
The large search model is an LLM that is customized to the search domain.
All IR tasks,
except the first-stage retrieval,
are formulated as text generation problems and are handled by a single large search model.
Given a user query and potentially thousands of retrieved documents,
the large search model generates the various elements that constitute the SERP,
including the ranked document list, document snippets, direct answers, etc.
Natural language prompts serve as the interface to customize the behavior of the model.
Different tasks are specified by different prompt templates,
and the adoption of LLM also allows for performing new tasks that are not explicitly trained.
Additionally,
ongoing research on multi-modal LLMs ~\citep{alayrac2022flamingo,huang2023language}
also allows the modeling of full document contents (text, images, videos, layout, etc.),
rather than just the textual portion of the documents.

To make our envisioned large search model ready for production systems,
several new challenges must be addressed.
For instance,
the inference cost of LLMs remains prohibitively high for real-time applications due to the autoregressive nature of text generation.
Additionally,
efficient long context modeling without compromising quality is still an open problem.
Finally,
ensuring that the generated content conforms to responsible AI principles is crucial for the deployment of this framework.
Many of these challenges are not unique to the search domain,
but are shared by other applications and have received extensive attention from the research community.

To empirically validate our approach,
we instantiate a simplified version of the large search model with the open-source LLaMA model ~\citep{Touvron2023LLaMAOA},
and conduct some preliminary experiments on joint listwise ranking and answer generation tasks.
The results show that our trained model is capable of achieving competitive performance compared to strong baselines.
To fully unleash the potential of large search models,
we call for further research to establish a benchmark setting and develop new methods
to tackle the challenges from the aspects of model architecture, training, inference, etc.

\section{Related Work}

\subsection{Neural Information Retrieval}
The field of information retrieval has undergone a paradigm shift in recent years,
transitioning from traditional lexical term-based models to neural models.
Neural information retrieval models ~\citep{mitra2018introduction} aim to overcome the limitations of traditional models,
such as the vocabulary mismatch problem and the lack of deep semantic understanding.
These models have demonstrated promising results in various applications,
including document retrieval ~\citep{xiong2020approximate}, document re-ranking ~\citep{nogueira2019multi},
query generation ~\citep{nogueira2019document}, and question answering ~\citep{Karpukhin2020DensePR}, etc.
Pre-trained Transformer-based language models ~\citep{Devlin2019BERTPO,Reimers2019SentenceBERTSE}
are widely used as the backbone of neural IR models,
and have been shown to be effective on several benchmark datasets ~\citep{nguyen2016ms,Kwiatkowski2019NaturalQA}.
Despite their success,
~\citet{Thakur2021BEIRAH} pointed out that neural IR models still underperformed the BM25 algorithm in out-of-domain scenarios.
Studies by ~\citet{Wang2022TextEB} show that large-scale contrastive pre-training can help to improve the generalization of neural models.
Combining the expressive power of neural models with
the lexical matching ability of term-based models ~\citep{Lin2021APC,Gao2021COILRE}
can further improve the robustness and interpretability of search systems.

\subsection{Retrieval-Augmented Generation}
Retrieval-augmented generation (RAG) endows language models with the capability to retrieve relevant information
from external knowledge sources ~\citep{ram2023context,lewis2020retrieval,shi2023replug},
and thus can improve the factuality and informativeness of the generated texts.
RAG also offers a natural approach to cite information sources ~\citep{nakano2021webgpt,Liu2023WebGLMTA},
which enhances the verifiability of the information presented.
The retrieved information can be incorporated into the generation process in different ways,
such as input concatenation ~\citep{ram2023context},
attention-based fusion ~\citep{borgeaud2022improving},
or output probability interpolation ~\citep{khandelwal2019generalization}.
Furthermore,
~\citet{lewis2020retrieval,zhong2022training} proposed
to jointly train the retrieval and generation components to promote their co-adaptation.
Nevertheless,
it is still unclear what is the optimal way to train RAG models,
and the degree to which the retrieved information is utilized in the generation process requires further investigation.

\subsection{Large Language Models}
Large language models (LLMs),
particularly decoder-only models such as LLaMA ~\citep{Touvron2023LLaMAOA} and GPT-4 ~\citep{OpenAI2023GPT4TR},
have shown great potential as general-purpose interfaces for natural language understanding and generation.
LLMs exhibit several attractive emergent behaviors,
including few-shot in-context learning, instruction following ~\citep{ouyang2022training},
and chain-of-thought reasoning ~\citep{wei2022chain} etc.
In the context of information retrieval,
LLMs have been utilized for pseudo-query generation ~\citep{dai2022promptagator},
pseudo-document generation~\citep{gao2022precise,wang2023query2doc}, long-form question answering ~\citep{nakano2021webgpt},
and ranking through prompting ~\citep{sun2023chatgpt}.
While existing work mainly focuses on the application of LLMs to specific search tasks,
we propose a conceptual framework to unify various search tasks with one LLM.

\section{Large Search Model}

\begin{figure*}[ht]
\centering
\includegraphics[width=1.0\textwidth]{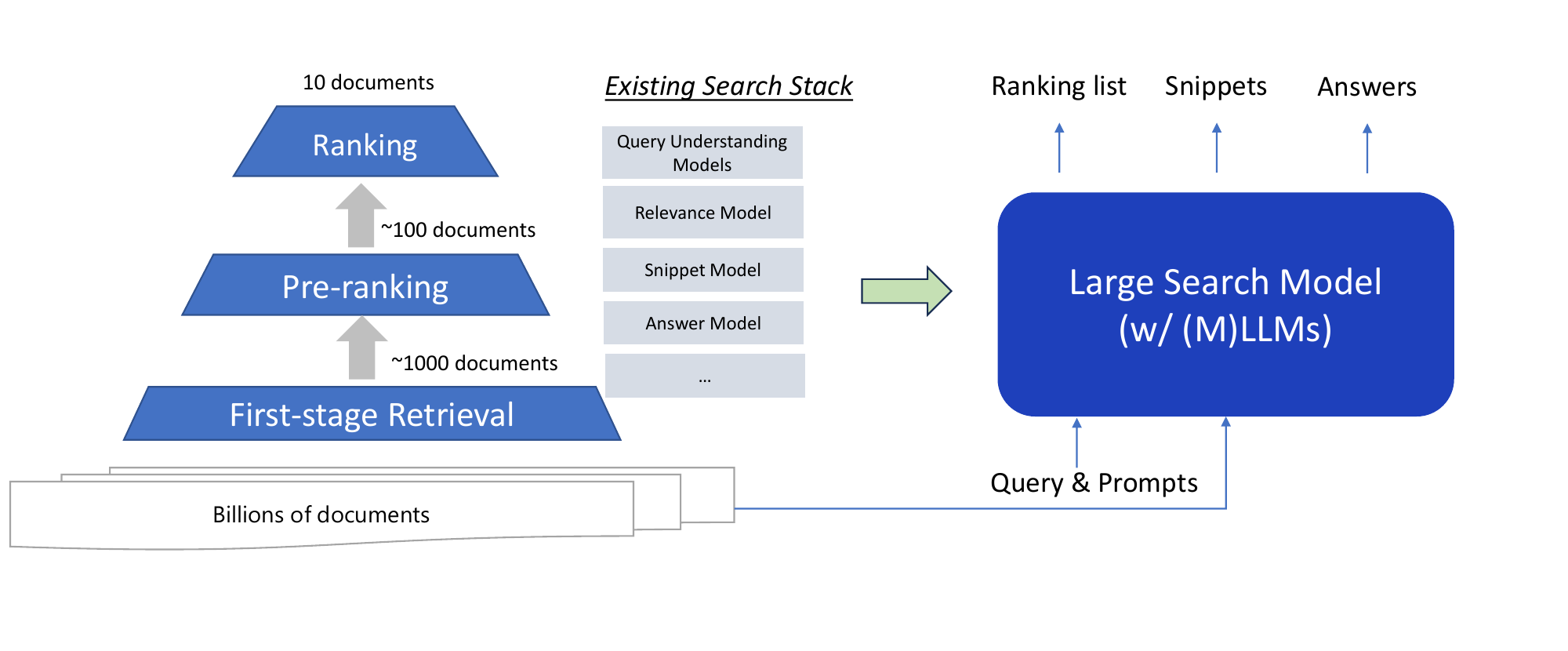}
\caption{Comparison of the conventional search stack and our proposed large search model.
The conventional search stack comprises a cascading retrieval and ranking pipeline,
along with many other components to generate the Search Engine Result Page (SERP).
In contrast,
our proposed framework employs a unified modeling approach,
where prompts are utilized to customize the large search model for diverse search tasks.
It is worth mentioning that the figure presented herein is for illustrative purposes only
and does not correspond to any specific implementation of modern search engines.
MLLM stands for Multi-modal Large Language Models.}
\label{fig:lsm_arch}
\end{figure*}

\subsection{Unified Modeling of Information Retrieval Tasks}

\begin{table}[ht]
\centering
\scalebox{0.9}{\begin{tabular}{lccccc}
\hline
                         & \# Models &  Fine-tuning & Latency & Cost & Task Generalization \\ \hline
Fine-tuned Enc-Dec / Enc &  Many & \cmark & Low & Low & \xmark  \\
Prompting GPT-4          &  1 & \xmark &  High  & High  &  \cmark \\ \hline
Large Search Model (ours)  &  1 & \cmark & Medium  & Medium  &  \cmark \\ \hline
\end{tabular}}
\caption{Comparison between three paradigms from different dimensions.
\emph{``Fine-tuned Enc-Dec / Enc''} refers to fine-tuning small encoder-decoder models or encoder-only models,
which is the prevalent approach at present.
Our proposed framework can be fine-tuned on IR tasks,
necessitating a smaller model size to achieve comparable performance to GPT-4.
This leads to a reduction in inference latency and cost.
However, we acknowledge that this argument requires further empirical validation.}
\label{tab:comparison_lsm_gpt4}
\end{table}

We define the concept of \emph{large search model} as a customized LLM (which may be multi-modal) that
can perform various IR tasks robustly through natural language prompting,
as depicted in Figure ~\ref{fig:lsm_arch}.
The technical implementation is not confined to any particular architecture or training objective,
and is not the central theme of this paper.

Compared to the current mainstream production systems that fine-tune separate small-size models such as T5 or BERT for each task,
the primary advantage of large search model lies in its ability to enhance task performance through a unified modeling approach.
This anticipated improvement in generalization capability comes from the increased model capacity of LLMs
and the possibility of leveraging the knowledge from other related tasks.
Of course,
this comes at the expense of increased latency and inference costs,
which is a major challenge for large-scale deployment of LLMs.
In contrast to prompting state-of-the-art proprietary models like GPT-4,
our proposed framework offers the flexibility to customize the model to desired search scenarios by fine-tuning on domain-specific data,
which are often abundant in commercial search engines.
A comprehensive comparison is presented in Table ~\ref{tab:comparison_lsm_gpt4}.

Modern search systems rely on collaborative efforts from many components to generate the result page.
Large search model can potentially replace many of these components,
and may even make some of them obsolete.
Here we provide a non-exhaustive list of search tasks in different phases of search systems that can benefit from our method:
\begin{itemize}
    \item \textbf{Online Serving } Responding to real-time user queries requires \emph{ranking over first-stage retrieval results}, \emph{answer generation}, \emph{snippet generation} and \emph{query suggestion} etc.
    \item \textbf{Data Augmentation } Model-based \emph{query generation} and \emph{relevance labeling} can be used to augment the training data for ranking models.
    \item \textbf{Indexing } It is an offline procedure to build a queryable document corpus,
    which requires \emph{content extraction}, \emph{term weighting}, and \emph{document expansion} etc.
    \item \textbf{Human Evaluation } Many queries are long-tailed or ambiguous, thus are difficult to evaluate for human raters.
    Automatic \emph{query intent generation} can help lower the cognitive burden of human raters and improve evaluation quality and efficiency.
\end{itemize}

One notable exception is the first-stage retrieval component,
which is currently implemented with embedding-based or lexical-based retrieval methods in our framework.
In contrast,
~\citet{metzler2021rethinking,tay2022transformer} propose a paradigm where the retrieval corpus is implicitly encoded in the model parameters,
thereby eliminating the need for a separate index.
However,
we argue that this paradigm  is still in its infancy and has serious limitations to be addressed,
such as how to scale to large corpus ~\citep{pradeep2023does} and how to dynamically update the corpus ~\citep{tay2022transformer}.

\subsection{Customization through Prompting}

By formulating various IR tasks as text-to-text generation problems,
we can build a unified large search model and specify which task to perform through natural language prompting.
This idea was initially popularized by T5 ~\citep{raffel2020exploring} and subsequently developed by decoder-only LLMs such as GPT-3.
LLMs can be directed to generate text in two primary ways:
\begin{itemize}
    \item \textbf{In-context Learning} The task is implicitly specified through a set of in-context examples,
    with each example comprising an input-output text pair.
    Language models need to learn the pattern of the input-output pairs and generalize to new unseen examples.
    This approach requires access to some labeled data and typically results in lengthy prompts.
    \item \textbf{Instruction} The task is specified by natural language instructions without any labeled data.
    This paradigm is more flexible and user-friendly.
    LLMs after instruction tuning ~\citep{ouyang2022training} can often generalize to new instructions that are not seen during training.
\end{itemize}

In Table ~\ref{tab:customization_prompts},
we give some examples to demonstrate how to instantiate different tasks within one model by customizing the prompt.
Different from the traditional multitask learning approach,
this paradigm does not add any task-specific parameters and
enables zero-shot generalization to new tasks during inference.

\begin{table*}[]
\centering
\scalebox{0.85}{\begin{tabular}{p{0.1\linewidth} | p{1.0\linewidth}}
\hline
Task & Document Ranking \\ \hline
Example &\begin{tabular}[c]{p{0.99\linewidth}}Query: what is wifi vs Bluetooth\\ \#\#\# Documents\\ {[}1{]} Takeaway: Bluetooth and Wi-Fi: Find out what separates these two wireless technologies. Source: Flickr/Dana Spiegel. Bluetooth and Wi-Fi are both methods that provide wireless communication, but the difference between the two mainly stems from what they are\ldots \\ \ldots \textit{(truncated)} \\ \#\#\# End of documents\\ Rank the documents according to their relevance to the query.\\ \\ A: \textbf{2\textgreater{}4\textgreater{}1\textgreater{}3\textgreater{}6 \ldots} \end{tabular} \\ \hline \hline
Task & Question Answering \\ \hline
Example & \begin{tabular}[c]{p{0.99\linewidth}}Q: what is hearing loss\\ \#\#\# Documents\\ {[}1{]} Hearing loss, also known as hard of hearing, anacusis or hearing impairment, is a partial or total inability to hear. It may occur in one or both ears\ldots \\ {[}2{]} Hearing loss, or hearing impairment, is a partial or total loss of the ability to hear and can affect one or both ears\ldots \\ \ldots\textit{(truncated)} \\ \#\#\# End of documents\\ Write an answer to the query based on the given documents.\\ \\ A: \textbf{Hearing loss, also known as hearing impairment, is a partial or total inability to hear in one or both ears. It can be caused by factors such as noise, genetics, aging, infections, or trauma}\ldots \textit{(truncated)} \end{tabular} \\ \hline \hline
Task & Snippet Generation \\ \hline
Example & \begin{tabular}[c]{p{0.99\linewidth}} Q: yoga therapy benefits \\ D: Mindful Yoga for Trauma Training for Yoga Teachers | 15-Hour Program For many of us, the practices of yoga have changed our lives. That statement may seem dramatic\ldots \textit{(truncated)} \\ Generate a short snippet based on the given document to answer the given query. \\ \\ A: \textbf{Yoga therapy benefits individuals coping with trauma-related difficulties, particularly veterans with Post Traumatic Stress. Mindful Yoga Therapy comprises five practices that provide symptom relief and support daily life skills. Although not a cure, yoga aids the healing journey and stimulates the overall healing process.} \end{tabular} \\ \hline \hline
Task & Query Intent Generation \\ \hline
Example & \begin{tabular}[c]{p{0.99\linewidth}}If a user searches for ``what county is aurora mo in'', what is this user's intent? \\ \\ \textbf{The user's intent is to find out which county the city of Aurora, Missouri is located in. They are looking for the name of the county that encompasses the city.} \end{tabular} \\ \hline \hline
Task & Query Suggestion \\ \hline
Example & \begin{tabular}[c]{p{0.99\linewidth}}A user searches for the query ``what glass made of'', what other related queries that might be of interest to this user? \\ \\ \textbf{How is glass made?} / \textbf{Different types of glass and their uses} / \textbf{The history of glassmaking} / \textbf{The science behind glass transparency} \end{tabular} \\ \hline \hline
\end{tabular}}
\caption{Instantiation of different information retrieval tasks through prompting.
The texts in bold are the expected outputs of the corresponding tasks.
This list is not exhaustive and only serves as a demonstration of the flexibility of our proposed framework.
Long texts are truncated for brevity.}
\label{tab:customization_prompts}
\end{table*}

\subsection{Long Context Modeling}

One of the shortcomings of current LLMs is that
they can only process a limited amount of context at a time.
During training,
the input texts are typically split into chunks of fixed length
to mitigate the quadratic computational complexity of self-attention.
Extending the effective context length requires either continual fine-tuning on longer texts
or designing special position interpolation strategies ~\citep{chen2023extending}.
Another concern over inference with long context is the increased memory consumption,
due to the need to cache key-value states in autoregressive decoding.
Techniques like multi-query attention ~\citep{shazeer2019fast} and grouped-query attention ~\citep{ainslie2023gqa}
are commonly employed to reduce the inference memory footprint at the cost of model quality.
Although substantial progress has been made in extending context length,
from the $512$ tokens for BERT to $4k$ for LLaMA-2, and $32k$ for GPT-4 ~\citep{OpenAI2023GPT4TR},
they are still often inadequate for real-world application scenarios.
Furthermore,
recent research ~\citep{Liu2023LostIT} has raised doubts about whether LLMs can effectively utilize long contexts as they claim.

Long context modeling is an essential capability requirement for our envisioned large search model.
Here we list several search-related tasks that can greatly benefit from such capability:
\newline

\noindent
\textbf{Long Document Retrieval and Ranking }
A significant portion of web documents are long-form texts,
such as news articles, legal documents, scientific papers, and code repositories, among others.
However,
most of the existing ranking and retrieval models only support short text inputs.
For example,
both the Sentence-BERT ~\citep{Reimers2019SentenceBERTSE} and E5 ~\citep{Wang2022TextEB} family of models
only support inputs of up to $512$ tokens.
Although some heuristic methods,
such as \emph{FirstP} and \emph{MaxP} ~\citep{xiong2020approximate},
are proposed to bypass the difficulty of long document modeling,
their limitations are also obvious.
\newline

\noindent
\textbf{Retrieval-Augmented Generation (RAG) }
LLMs are known to hallucinate facts especially when the generation requires long-tailed knowledge,
and they are not aware of the latest events or information that are simply not publicly available.
Retrieval augmentation is a widely adopted technique to mitigate this problem.
However,
when generation requires conditioning on many retrieved documents,
the context length can easily exceed the maximum input length of LLMs,
even if each document can individually fit within the length limit.
Take the task of open-domain question answering as an example,
~\citet{izacard2021leveraging} found that aggregating information from $100$ passages achieves the best performance.
If concatenating all passages into a single input,
the resulting length can be well beyond the maximum context length of many LLMs.
\newline

\noindent
\textbf{Conversational Search }
Conversational interfaces, popularized by ChatGPT,
have been integrated into several mainstream search engines,
such as New Bing ~\footnote{\url{https://www.bing.com/new}} and Google Bard ~\footnote{\url{https://bard.google.com/}}.
They provide a more natural way for users to interact with search engines through multi-turn conversations.
As the conversation progresses,
the conversation history can grow to a considerable length.
Ideally,
a conversational search system should be able to utilize the entire conversation history to model the user's intent and generate the most relevant responses.
This requires LLMs to be capable of processing long contexts.

\subsection{Multi-modal Large Search Model}

Multi-modal contents other than plain texts are ubiquitous on the web,
including images, videos, audio, and other rich media formats.
Incorporating such information into the search model can significantly improve the quality of search results
and enable new search experiences.
Users can submit queries in mixed modalities,
and the search engine will render the results in the most appropriate format.
Existing commercial systems have some but limited support for multi-modal search in the form of independent modules,
such as stand-alone image search, video search, and music search tabs.

Here we argue that the development of multi-modal foundation models ~\citep{Radford2021LearningTV,alayrac2022flamingo,huang2023language}
could bring tremendous opportunities to the next generation of search engines.
Large multi-modal foundation models can provide a deep understanding of web content,
regardless of the modalities they exist in,
contrasting to the current systems that almost exclusively rely on texts.
They can also be used to synthesize multi-modal contents that better serve the user's information needs
and provide a more immersive experience.
Products like \emph{Bing Image Creator} ~\footnote{\url{https://www.bing.com/create}} powered by DALL-E 2 ~\citep{Ramesh2022HierarchicalTI}
are already taking the first step in this direction.
Developing larger and more robust multi-modal foundation models is a fast-evolving research area,
and we expect that they will unlock numerous possibilities for web content understanding and generation.

\subsection{Practical Considerations for Deployment}

Deploying LLMs in a real-world search system is a challenging task that requires careful planning and optimization.
They need to be scaled up to handle tens of thousands of queries per second from users around the world under strict latency constraints
while providing more accurate and useful information compared to the existing search stack.
Some of the practical considerations for our envisioned system are as follows:
\newline

\noindent
\textbf{Inference Efficiency }
LLMs typically have billions of parameters and require huge amounts of computational and memory resources to run.
Besides relying on the development of more powerful hardware,
common techniques that can be used to improve the efficiency include model compression, quantization, pruning, and kernel fusion ~\citep{han2015deep,dao2022flashattention}.
Sparse Mixture-of-Experts (MoE) architecture ~\citep{fedus2022switch} is also a promising direction
to increase the model capacity while keeping the inference cost manageable.
However,
most open-source LLMs still follow the dense Transformer architecture.
Another line of research seeks to speed up the autoregressive decoding process by attempting to generate multiple tokens in each inference step.
\emph{Speculative decoding} ~\citep{leviathan2023fast} adopts a small language model to generate a few candidate tokens,
which are then verified with the target LLM,
while the \emph{inference with reference} ~\citep{yang2023inference} method guesses the future tokens based on the available context.
Additionally,
caching mechanisms can be utilized to avoid online decoding for frequently seen queries.
\newline

\noindent
\textbf{Hallucination }
LLMs may hallucinate texts that are incorrect, nonsensical, or inconsistent with the given information.
For example,
LLMs may generate a factually inaccurate statement about a historical event,
or a fictional entity that does not exist in the real world.
Hallucinations can have serious consequences for users who rely on LLMs for accurate information, education, or decision making.
Studies ~\citep{nakano2021webgpt,OpenAI2023GPT4TR} suggest that hallucinations can be reduced by retrieval augmentation and further scaling up of LLMs,
but it remains an unsolved open problem.
\newline

\noindent
\textbf{Alignment }
Language model alignment ~\citep{kenton2021alignment,ouyang2022training} has multi-faceted meanings.
Here, we mainly focus on the alignment between LLMs and human values.
As the training data of LLMs often contains undesirable content,
LLMs without alignment can generate outputs that are offensive, biased, or even harmful to users.
Even for models after alignment,
carefully designed adversarial prompts can still mislead the model to generate inappropriate outputs.
To deploy LLMs in a useful and responsible way,
comprehensive measures must be taken as safeguards,
including data filtering, content moderation, and red-team testing ~\citep{ganguli2022red}, etc.


\section{Proof-of-Concept Experiments}

In this section,
we present some proof-of-concept experiments to showcase the potential of our proposed framework.
Specifically,
we fine-tune the LLaMA-7B model ~\citep{Touvron2023LLaMAOA} on the MS MARCO passage ranking dataset ~\citep{nguyen2016ms}
for two tasks: listwise ranking and retrieval-augmented answer generation.
Both tasks are framed as text generation problems,
and we only compute the cross entropy loss on the target tokens.
The prompt templates are taken from Table ~\ref{tab:customization_prompts}.
Since the original dataset does not provide annotations for listwise ranking,
we use the re-ranker score from ~\citet{Wang2022SimLMPW} as ranking labels
while always placing the human-annotated positive passage in first place during training.
For the answer generation task,
we collect the outputs for $40k$ queries from \emph{gpt-35-turbo} ~\footnote{\url{https://oai.azure.com/portal}} as the ground truth answers.
The input passages are the top-100 retrieval results from an off-the-shelf dense retriever E5$_\text{large-v2}$ ~\citep{Wang2022TextEB}.
We utilize the linear positional interpolation method ~\citep{chen2023extending} and skip encodings ~\citep{zhu2023pose} to extend the context length from $2k$ to $16k$.
The model is trained for one epoch with a batch size of 128 and a learning rate of $10^{-5}$.

\begin{table}[ht]
\centering
\scalebox{0.9}{\begin{tabular}{lccc}
\hline
                                & MS MARCO & TREC DL 19 & TREC DL 20 \\ \hline
BM25                            &    18.4      &   51.2         &    47.7        \\
ANCE ~\citep{xiong2020approximate}        &   33.0       &   64.5  &  64.6   \\
E5$_\text{large-v2}$ ~\citep{Wang2022TextEB}   &   38.4   &  70.9  & 72.1   \\ \hline
Ours (Listwise rank + LLaMA$_\text{7b}$) &   \textbf{41.7}   &   \textbf{72.9}   &  \textbf{74.0}   \\ \hline
\end{tabular}}
\caption{Listwise ranking results on the MS MARCO passage ranking and TREC Deep Learning tracks.
Our model takes the top-100 passages retrieved by E5$_\text{large-v2}$ as input
and outputs top-5 passages with greedy decoding.
Consistent with previous work,
we report MRR@10 for MS MARCO and NDCG@10 for TREC DL 19/20.}
\label{tab:listwise_rank_results}
\end{table}

\begin{table*}[ht]
\centering
\scalebox{0.8}{\begin{tabular}{p{0.2\linewidth} | p{1.0\linewidth}}
\hline
Query & rsa definition key \\ \hline
Retrieved contexts & \begin{tabular}[c]{p{0.99\linewidth}} {[}1{]} RSA is a cryptosystem for public-key encryption, and is widely used for securing sensitive data, particularly when being sent over an insecure network such as the Internet. \\ {[}2{]} RSA algorithm RSA (algorithm) RSA is an algorithm used by modern computers to encrypt and decrypt messages. It is an asymmetric cryptographic algorithm. Asymmetric means that there are two different keys. This is also called public key cryptography, because one of them can be given to everyone. The other key must be kept private. \\ {[}3{]} RSA is a cryptosystem for public-key encryption, and is widely used for securing sensitive data, particularly when being sent over an insecure network such as the Internet. Download this guide to become a Web App Security superhero. \\ \ldots \textit{(97 more, truncated)} \end{tabular} \\ \hline
Model generation & RSA is a public-key cryptosystem that uses two different keys, one public and one private. The public key is used for encryption, while the private key is used for decryption. The asymmetry in RSA is based on the difficulty of factoring the product of two large prime numbers, which is the factoring problem. The public key is used to encrypt messages, while the private key is used to decrypt them. The RSA algorithm is based on the difficulty in factoring very large numbers, and deducing an RSA key takes a huge amount of time and processing power. RSA is widely used for securing sensitive data, particularly when being sent over an insecure network such as the Internet. \\ \hline
GPT-4 Score (1-10) & 9 \\ \hline
\end{tabular}}
\caption{An example for answer generation.
Our model generates an answer for the given query ``rsa definition key'' conditioned on the top-100 retrieved contexts.}
\label{tab:rag_example}
\end{table*}

Table ~\ref{tab:listwise_rank_results} indicates that our model outperforms BM25 sparse retrieval and multiple strong dense retrievers,
thus showcasing its potential for listwise ranking.
For answer generation,
we assess the quality of the generated answers with GPT-4 on a scale of 1 to 10,
where higher scores denote better quality.
On the TREC DL 2019 queries,
our model achieves an average score of $8.4$,
while its teacher model \emph{gpt-35-turbo} gets $8.9$.
Upon manual inspection,
we find that most generated answers are fluent and relevant to the query,
but some of them are not as comprehensive as the ground truth answers and could contain repetitions.
Table ~\ref{tab:rag_example} presents an example of the generated answers.
Further experiments on additional tasks and datasets will be left for future endeavors.

\section{Conclusion}

This paper introduces the \emph{large search model} framework to redefine the technique stack of search systems in the era of LLMs.
We argue that the unique characteristics of LLMs permit the adoption of a unified modeling approach
for various IR tasks and offer improved generalization ability,
instead of fine-tuning and deploying numerous task-specific small encoder-decoder or encoder-only models.
Along with the great potential of this framework,
we also discuss several emerging challenges that necessitate further research,
such as high inference cost, long context modeling, and the potential risks of misalignment, etc.
To demonstrate the feasibility of our framework,
proof-of-concept experiments are conducted,
although a larger-scale evaluation is required for a more comprehensive assessment.

While modern search engines have been instrumental in democratizing access to information,
building a robust search system demands an enormous amount of engineering effort spanning multiple components,
and the search results are still not satisfactory in many cases.
We believe that the ongoing development of LLMs will bring a new wave of innovation to the field of information retrieval,
and we hope that our work can inspire further research in this direction.

\section*{Acknowledgments}
We would like to thank many members from the Microsoft Bing team for their valuable feedback and suggestions.
We also thank the SIGIR Forum for providing the platform to share opinions and ideas.

\bibliography{sigirforum}
\end{document}